\documentclass[12pt]{article}
\usepackage{epsfig}

\usepackage{graphicx}
\usepackage{cancel}
\usepackage{amssymb,amsmath}

\def\harr#1#2{\smash{\mathop{\hbox to .3in{\rightarrowfill}}
 \limits^{\scriptstyle#1}_{\scriptstyle#2}}}

\def\s2{\frac{1}{\sqrt2}}

\def\beqa{\begin{eqnarray}}
\def\eeqa{\end{eqnarray}}

\def\Dsl{\,\raise.15ex\hbox{/}\mkern-13.5mu D} 
\def\d3{d^3}

\newcommand{\be}{\begin{equation}}
\newcommand{\ee}{\end{equation}}
\newcommand{\beq}{\begin{eqnarray}}
\newcommand{\eeq}{\end{eqnarray}}
\newcommand{\m}{\mu}
\newcommand{\n}{\nu}
\newcommand{\f}{\frac}
\newcommand{\no}{\nonumber}
\newcommand{\p}{\partial}

\newcommand{\ove}{\overline}
\newcommand{\bz}{\overline{z}}

\newcommand{\lc}{\left[}
\newcommand{\rc}{\right]}

\begin{document}

\begin{center}
\Large{\bf Self-dual gravity via Hitchin's equations} \vspace{1cm}

\large  Erick Chac\'on\footnote{e-mail address: {\tt
echacon@fis.cinvestav.mx}},
           Hugo Garc\'{\i}a-Compe\'an\footnote{e-mail address: {\tt compean@fis.cinvestav.mx}}\\

\vspace{0.5cm}
{Departamento de F\'{\i}sica,}\\
Centro de Investigaci\'{o}n y de Estudios Avanzados
del Instituto Polit\'{e}cnico Nacional\\
{P.O. Box 14-740, CP. 07000, M\'exico D.F., M\'exico.}\\

\vspace*{1cm}

\end{center}

\begin{abstract}
In this work half-flat metrics are obtained from Hitchin's
equations. The SU$(\infty)$ Hitchin's equations are obtained and as
a consequence of them, the Husain-Park equation is found.
Considering that the gauge group is SU$(2)$, some solutions
associated to Liouville, sinh-Gordon and Painlev\'e III equations
are taken and, through Moyal deformations, solutions of the
SU$(\infty)$ Hitchin's equations are obtained. From these solutions,
hamiltonian vector fields are determined, which in turn are used to
construct the half-flat metrics. Following an approach of Dunajski,
Mason and Woodhouse, it is also possible to construct half-flat
metrics on ${\cal M} \times\mathbb{CP}^{1}$.

\vskip 3truecm

\noindent\leftline{December 21, 2018}
\end{abstract}

\bigskip
\newpage
\section{Introduction}

Several years ago Ward in Ref. \cite{Ward85,Wardconjecture}
conjectured that all non-linear integrable systems in field theory
in lower dimensions than four might be obtained by dimensional
reduction from four-dimensional self-dual Yang-Mills (SDYM)
equations using finite gauge group. This is called Ward's conjecture
and it can be generalized to include all sl$(N,\mathbb{C})$ algebras
as well as infinite dimensional algebras of hamiltonian vector
fields. This conjecture has been a guiding principle in a great deal
of work in integrable systems done since then (for some complete
reviews see, \cite{Mason:1991rf,Dunajski:2010zz}).

In a remarkable article Ashtekar, Jacobson and Smolin (AJS)
\cite{Ashtekar:1987qx}, reduced the self-dual gravity (SDG)
equations to a set of three volume-preserving vector fields
satisfying a Nahm-like equation. Later Mason and Newman
\cite{Mason89}, shown that starting from the SDYM equations with
gauge group $G$ in four dimensions, it is possible to obtain
self-dual gravity equations (in the AJS version) in vacuum space.
This is achieved by assuming that the vector potentials are
independent of all the spacetime coordinates and the Lie algebra
generators of $G$ are replaced by the volume preserving hamiltonian
vector fields $X_f$ of the underlying four manifold $M$, where $f$
is the corresponding hamiltonian function. Thus the Lie brackets of
the Lie algebra of $G$ are changed by the Lie bracket of hamiltonian
vector fields. At the level of hamiltonian functions, the Lie
brackets are replaced by Poisson brackets. The relationship of the
obtained AJS equations to the Pleba\'nski heavenly equations
\cite{Plebanski75} was later obtained in Ref.
\cite{Chakravarty:1991bt}. In this same direction Dunajski made a
generalization of AJS equations to the hyper-Hermitian case
\cite{Dunajski:1998nj}.

It is also possible to obtain SDG from SDYM if a two dimensional
reduction is performed with gauge group $G=$SDiff$(\Sigma)$
\cite{Mason2, Ward90,Compean94}. Mason observed that
SL$(N,\mathbb{C})$ is a subgroup of SDiff$(\Sigma)$ only for $N=2$
\cite{Mason2}.  It is know that solutions of the SDYM equations with
two-dimensional reductions are singular if the solution is real and
is regular if the solution is complex \cite{Lohe77,Saclioglu84}.

Moreover, starting from two-dimensional conformal field theories it
is also possible to obtain SDG in four dimensions by promoting the
gauge group $G$ of the two-dimensional theory to be the
area-preserving diffeomorphism group, SDiff$(\Sigma)$. Some of these
examples were explicitly given by Park \cite{Park90}, who showed
that SDG arises from two-dimensional SU$(N)$ non-linear sigma model
with Wess-Zumino term in the large-$N$ limit. Later Park extended
this work and shown that it is possible to obtain SDYM equations on
a self-dual space if the gauge group SDiff$(\Sigma)$ is extended to
an arbitrary Lie algebra \cite{Park92}. Later Husain \cite{Husain94}
showed how an alternative to the Pleba\'{n}ski heavenly equation
\cite{Plebanski75} could be derived from the two-dimensional chiral
model.  Following this line of thought the non-linear graviton was
constructed from chiral fields in Ref. \cite{Przanowski:1999js}.

Another way to find solutions to the heavenly equations is by using
the Moyal deformation quantization method \cite{Zachos:2001ux}. The
Moyal deformation of the heavenly equations was considered in
\cite{Strachan:1992em,Takasaki:1992jf,Takasaki:1993my,
Plebanski:1995gk,Castro:1997hr}.  Using this Moyal deformation on
the two-dimensional chiral model with the Lie bracket replaced by
the Moyal one, it is possible to obtain solutions of the Husain-Park
heavenly equation \cite{Compean}. More generally, starting from the
SDYM equations a six-dimensional master equation can be obtained
through the deformation quantization procedure
\cite{Plebanski:1996np}. From this master equation, by dimensional
reduction, it is possible to find various integrable systems in four
dimensions, one of them is the Husain-Park heavenly equation. The
most general case of the SDYM fields in a SDG background and its
Moyal deformation was considered by Forma\'nski and Przanowski in
Refs. \cite{Formanski:2004dd,Formanski:2005wt}. For other
applications, see \cite{GarciaCompean:1996np,GarciaCompean:2009cg}.
In other contexts, the self-dual sector of gravity has been also
explored as related to the $N=2$ strings
\cite{Ooguri:1990ww,Ooguri:1991fp} and to the double copy conjecture
\cite{Monteiro:2011pc,Monteiro:2014cda,Berman:2018hwd,Luna:2018dpt}.
Furthermore, Monteiro and O'Connell in \cite{Monteiro:2011pc} were
able to identify an algebra underlying the color- kinematics duality
with that of area preserving diffeomorphisms.

Very recently a great deal of activity on Hitchin's equations, and
in general the theory of Higgs bundles, has been taken place and
their impact in physics and mathematics is strongly discussed.
Hitchin's equations \cite{Hitchin87} were originally obtained as a
two-dimensional reduction from the Euclidean SDYM equations in four
dimensions. He obtained a set of equations involving two-dimensional
gauge fields coupled to one-form section of a holomorphic vector
bundle over a Riemann surface $\Sigma$ called the Higgs bundle. Thus
these equations are characterized by the pairs $\{(E, \Phi)\}$,
where $E$ is a holomorphic bundle over $\Sigma$ and $\Phi$ is a
section of the bundle $\Omega^1 \otimes E$ such that $\Phi \wedge
\Phi =0$. The moduli space of Higgs bundles has very interesting
properties. For instance, in the context of Riemann surfaces, it
describe an integrable system \cite{Hitchin:1987mz,drinfeld}.
Further generalizations for holomorphic bundles over general
K\"ahler manifolds was considered by Simpson in Ref. \cite{simpson}.
Recently, a four-dimensional version of the Hitchin's system called
the Kapustin-Witten (KW) equations was obtained also in the context
of the physical approach to the geometric Langlands problem
\cite{Kapustin:2006pk}. In this approach, the KW equations were
obtained from the dimensional reduction of a topological Yang-Mills
theory in six dimensions. Moreover a great deal of work has been
done in this context, mainly in its relation to Khovanov homology
(for some reviews, see for instance, \cite{khovanov}). Another
interesting spinoff of Hitchin's equations in mathematics is its
relationship with the wall-crossing formula of Kontsevich-Soibelman
(see for instance \cite{Gaiotto:2009hg}).

The relationship between Hitchin's equations and gravity has been
few investigated. For instance, Ueno obtained the Hitchin's
equations starting from the SDG \cite{Ueno96}. Later Etesi gives a
AdS/CFT correspondence between classical $(2+1)$-dimensional vacuum
gravity on $\Sigma\times\mathbb{R}$ and SO$(3)$ Hitchin's equations
on the space-like past boundary $\Sigma$, a compact, oriented
Riemann surface of genus greater than $1$ \cite{Gabor07}. Moreover,
Calderbank showed that if we take Diff$({\bf S}^{1})$  to be the
gauge group in the Hitchin's equations then it is obtained the
hyperCR Einstein-Weyl equation. Otherwise if we take the gauge group
to be SDiff$(\Sigma)$ then we obtain an Euclidean analogue of
Pleba\'nski's heavenly equations \cite{Calderbank14}.  KW equations
were recently considered in the context of deformation quantization
\cite{Cardona:2017ebi}. In this paper the SU$(\infty)$ KW equations
arise in a natural way and the question of the possible relation of
this equation to gravity was raised.

In the present paper we will continued studying how self-dual
gravity arises directly from the SU$(N)$ Hitchin's equations in the
large-$N$ limit and from finite group. We will find the relationship
between SU$(\infty)$ Hitchin's equations and Husain-Park equation,
also the corresponding self-dual metrics of three different family
of solutions to the Hitchin's equations, namely: Liouville,
sinh-Gordon and  Painlev\'e III models, will be also found.

The structure of the paper is as follows. In Section 2 we review the
Hitchin's equations and give the Lax pair formalism. Using SU$(2)$
as gauge group, we review three families of solutions of Hitchin's
equations which are expressed in terms of the Liouville, sinh-Gordon
and Painlev\'e III equations. In the case of the Liouville equation
there are a singular and regular solution depending if the gauge
fields are complex or real. The solutions given by sinh-Gordon and
Painlev\'e III equations are singular and regular respectively. In
section 3 we give the SU$(\infty)$ version of the Hitchin's
equations in terms of the hamiltonian functions and find a relation with Husain-Park equation. Using Moyal
deformation, given a solutions of the Hitchin's equations one gets
solutions of the SU$(\infty)$ version, we can construct explicit
forms of the hamiltonian functions and their corresponding
hamiltonian vectors fields. In section 4 we use the fact that
SU$(2)$ is a sub group of SU$(\infty)$ and we construct vector
fields from hamiltonian vector fields over $\mathbb{CP}^{1}$. Also,
in section 5 we construct half-flat metrics from the vector fields
obtained previously using the correspondence between Lax pair
formulation of the Hitchin's equation and the self-dual gravity. The
half-flat metrics obtained are the Husain's metric and other six
half-flat metrics that depend on the solutions of the Liouville,
sinh-Gordon and Painlev\'e equations, thus these metrics can be
singular or regular. Finally, in section 6 we give our conclusions
and some open questions are raised.

\section{Hitchin's equations and some solutions}

In this section we review the Hitchin's equations coming from
two-dimensional reduction of SDYM equations over $\mathbb{R}^{4}$,
the equations obtained are conformally invariant. Later we find the
Lax pair for Hitchin's equations from a two-dimensional reduction of
the Lax pair for SDYM equations. Finally in this section, we review
three families of solutions of the SU$(2)$ Hitchin's equations
coming from two ans\"atze. These solutions will depend on three
different differential equations: Liouville, sinh-Gordon and
Painlev\'e III equations. Later in this paper we use these solutions
to build hamiltonian vectors fields and half-flat metrics.

On a coordinate system $\{x_{i}\}$ ($i=1, \dots ,4$) of Euclidean
space $\mathbb{R}^4$, the self-dual Yang-Mills (SDYM) equations
$\star F = F$ can be expressed as
\begin{equation}
F_{12}=F_{34},\quad F_{13}=F_{42},\quad F_{14}=F_{23}, \label{SDYM}
\end{equation}
where $\star$ is the Hodge star operator. The components of the
curvature are given by $F_{ij}=[D_{i},D_{j}]$, where the brackets
are the Lie brackets and $D_{i}=\p_{i}+A_{i}$ is the covariant
derivative. Moreover $A= \sum_i A_i(x) dx^i$ is the connection
one-form associated to the curvature $F$ i.e. $F = dA + A \wedge A$.
If we consider that $A_{i}$ will be only functions of the
coordinates $x_{1}$ and $x_{2}$, then the SDYM equations
(\ref{SDYM}) are explicitly written as
\begin{gather}
\p_{1}A_{2}-\p_{2}A_{1}+[A_{1}, A_{2}]= [A_{3}, A_{4}], \no \label{SDYM1}\\
\p_{1}A_{3}+[A_{1},A_{3}]=-\p_{2}A_{4}+[A_{4},A_{2}], \no \label{SDYM2}\\
\p_{1}A_{4}+[A_{1},A_{4}]=\p_{2}A_{3}+[A_{2},A_{3}].\label{SDYM3}
\end{gather}
Now we define the following field combinations
$$
A_{\ove{z}}=\frac{1}{2}(A_{1}+iA_{2}), \ \ \ \ \ \ \ \ \ \ \ \ \
A_{z}=\frac{1}{2}(A_{1}-iA_{2}),
$$
\begin{equation}
\Phi=A_{3}-iA_{4}, \ \ \ \ \ \ \ \ \ \ \ \ \
\Phi^{\ast}=-A_{3}-iA_{4}, \label{Fun2}
\end{equation}
where we have taken the complex coordinate $z=x+iy$, $x_{1}=x$,
$x_{2}=y$. In terms of these fields and coordinates, equations from
(\ref{SDYM3}) can be rewritten as
\begin{gather}
F_{z\ove{z}}+\frac{1}{4}[\Phi,\Phi^{\ast}]=0,\label{HitchinA1}\\
D_{\ove{z}}\Phi=\p_{\ove{z}}\Phi+[A_{\ove{z}},\Phi]=0\label{HitchinA2}.
\end{gather}

These equations are the well known Hitchin's equations
\cite{Hitchin87}, $\Phi$ and $\Phi^{\ast}$ are the Higgs fields.
They are invariant under gauge transformation of the form
\begin{align}
A_{z}&\to g^{-1}A_{z}g+g^{-1}g,_{z}, & \Phi&\to g^{-1}\Phi g, \no \\
A_{\bz}&\to g^{-1}A_{\bz}g+g^{-1}g,_{\bz}, & \Phi^{\ast}&\to
g^{-1}\Phi^{\ast} g,
\end{align}
where $g=g(z,\bz)$ is an element of the gauge group $G$.
\subsection{Lax pair}

Now, the Lax pair for the Hitchin's system is obtained from a
two-dimensional reduction from the Lax pair for SDYM and the reduced
system is obtained by making a particular choose of the fields. The
SDYM equations (\ref{SDYM}) have a compatibility condition given by
the Lax pair \be [L,M]=0, \ee with
\begin{gather}
L=D_{1}-iD_{2}-\lambda (D_{3}-iD_{4}), \no \\
M=-D_{3}-iD_{4}-\lambda (D_{1}+iD_{2}),
\end{gather}
where $D_{i}=\p_{i}+A_{i}$. If now we perform the two-dimensional
reduction, we assume that connections will be functions independent
on the coordinates $x_{3},x_{4}$, then from the definitions in
(\ref{Fun2}), the Lax pair for the Hitchin's system are given by \be
L=D_{z}-\frac{\lambda}{2}\Phi, \quad
M=\frac{1}{2}\Phi^{\ast}-\lambda D_{\ove{z}}. \ee  Thus the
compatibility condition reduces to \be
\bigg[D_{z}-\frac{\lambda}{2}\Phi,
D_{\ove{z}}-\frac{\lambda^{-1}}{2}\Phi^{\ast}\bigg]=0, \ee where we
also have de equation
$D_{z}\Phi^{\ast}=\p_{z}\Phi^{\ast}+[A_{z},\Phi^{\ast}]=0$, which
can be obtained from the SDYM reduction equations.

In the case for a particular value of spectral parameter
$\lambda=1$, the compatibility condition has the form
\begin{equation}
\bigg[D_{z}-\frac{\Phi}{2}, D_{\bz}-\frac{\Phi^{\ast}}{2}\bigg]=0.
\end{equation}
This give us the single equation
\begin{equation}
\left(A_{\bz}-\frac{\Phi^{\ast}}{2}\right),_{z}-
\left(A_{z}-\frac{\Phi}{2}\right),_{\bz}+\left[A_{z}-\frac{\Phi}{2},
A_{\bz}-\frac{\Phi^{\ast}}{2}\right]=0. \label{LPD}
\end{equation}
Then there exist a gauge transformation in which we can take
$\Phi=2A_{z}$ and $\Phi^{\ast}=2A_{\bz}$. Substituting these values
into (\ref{HitchinA1}) and (\ref{HitchinA2}), and taking into
account also
$D_{z}\Phi^{\ast}=\p_{z}\Phi^{\ast}+[A_{z},\Phi^{\ast}]=0$, the
Hitchin's equations reduce to
\begin{equation}
A_{z},_{\bz}-[A_{z},A_{\bz}]=0, \quad A_{\bz},_{z}+A_{z},_{\bz}=0.
\label{Hitchinredu}
\end{equation}
Later these equations will be related with Husain-Park equation when
the gauge group is SU$(\infty)$.

\subsection{Some solutions}

In this subsection we review some solutions of Hitchin's equations,
with gauge group SU$(2)$, related to Liouville, sinh-Gordon and
Painlev\'e III equations. In the case of the Liouville equation it
is possible to obtain analytic solutions which will be singular or
regular depending whether the connections are real or complex and
are defined over ${\bf S}^{2}$. In the case of sinh-Gordon equation
we have an analytic solution but it is singular. For the Painlev\'e
III case, the corresponding solution is defined over
$\mathbb{R}^{2}$, it is regular but there is not an analytic
solution.

In Ref. \cite{Mosna07} Mosna and Jardim obtained solutions to Hitchin's equations given the following ansatz. For the gauge group SU$(2)$, they proposed the following form for the gauge connections
\begin{align}
A_{z}&=\frac{1}{2}(f_{1}-if_{2})\tau_{1}, & A_{\ove{z}}&=\frac{1}{2}(f_{1}+if_{2})\tau_{1}, \no \\
\Phi &= (i)^{n}g\tau_{2}-(i)^{n+1}h\tau_{3}, &
\Phi^{\ast}&=-(i)^{n}g\tau_{2}-(i)^{n+1}h\tau_{3},
\end{align}
where $\tau_{i}$ is the basis for the Lie algebra $\mathfrak{su}(2)$
and $n=0,1$. Hitchin's equations hold if the functions $f_1,f_2,g$
and $h$ satisfy the following relations
\begin{gather}
\p_{x}g = f_{2}h,\quad \p_{y}g =- f_{1}h, \no \\
\p_{x}h=f_{2}g,\quad \p_{y}h=-f_{1}g, \no \\
\p_{x}f_{2}-\p_{y}f_{1}=(i)^{2n}gh.
\end{gather}
We have an integration constant $\kappa$ given by \be
\kappa^{2}=g^{2}-h^{2}. \ee There are two independent cases
$\kappa=0$ and $\kappa\neq 0$. In the first case it gives rise to
the Liouville equation, which has singular and regular solutions,
depending if the solutions of the SDYM equations are real or
complex. In the second case,  the equation is the sinh-Gordon and it
determines a real (singular) solution of the SDYM equations.

\subsubsection{Liouville equation}

The solution of the Hitchin's equations corresponding to the
Liouville equation is defined on ${\bf S}^{2}$ and we will consider
the real and complex cases depending on the choice of $n=0$ or $n=1$
respectively. Thus we have $\kappa=0$ and $g=h$, the functions
$f_{1}$ and $f_{2}$ are then given as a function of $g$
\begin{gather}
f_{1}=-\p_{y} \ln g,\quad f_{2}=\p_{x} \ln g, \no \\
\p_{x}f_{2}-\p_{y}f_{1}=(i)^{2n}g^{2}.
\end{gather}
Then the function $g^{2}=\lambda$ is obtained from the Liouville
equation, depending on the choose of $n$, i.e. \be
\nabla^{2}(\ln\lambda)\mp 2\lambda =0. \label{Liouville} \ee Here
the top sign corresponds to $n=0$, while the bottom sign corresponds
to $n=1$ and $\nabla^{2}=\p_{x}^{2}+\p_{y}^{2}$. Equation
(\ref{Liouville}) can be rewritten in complex coordinates $z=x+iy$
as follows \be \p_{z}\p_{\ove{z}}(\ln\lambda)\mp\frac{\lambda}{2}=0.
\ee Then a solution of this equation has the general form \be
\lambda(z,\ove{z})=4\frac{\psi'(z)\eta'(\ove{z})}{[1\mp\psi(z)\eta(\ove{z})]^{2}},
\ee where $\psi$ and $\eta$ are arbitrary functions and $'$ stands
for derivative with respect to $z$. Real solutions can be found by
taking $\eta(\ove{z})=\ove{\psi}(z)$. If in addition we have
$\psi=z^{\nu}$, the function $g$ will be a function only depending
on $r=|z|$, and it is given by \be
g^{2}(r)=4\nu^{2}\frac{r^{2\nu-2}}{(1\mp r^{2\nu})^{2}}\label{g}.
\ee The gauge connection defined on the plane $x-y$ has the
following form \be A=A_{1}dx_{1}+A_{2}dx_{2}=r\p_{r}(\ln
g)\tau_{1}d\theta. \ee Taking into account the form of $g$ in
(\ref{g}), the connection can be written as \be A=\left (\nu-1\pm
2\nu\frac{r^{2\nu}}{1\mp r^{2\nu}}\right )\tau_{1}d\theta. \ee In
the case $\nu=1$ it is reduced to \be A=\pm 2\frac{r^{2}}{1\mp
r^{2}}\tau_{1}d\theta = M\tau_{1}, \ee which can be related with an
abelian magnetic monopole $M=\pm\frac{2r^{2}}{1\mp r^{2}}d\theta$.

With $g$ given by (\ref{g}), we obtain an explicit solution to the
Hitchin's equations \beq
A_{z}&=&-i\p_{z}(\ln g)\tau_{1}=-\frac{i}{2}\frac{e^{-i\theta}}{r}\left (\nu-1\pm2\nu\frac{r^{2\nu}}{1\mp r^{2\nu}}\right )\tau_{1}, \no \label{Mosna1}\\
A_{\ove{z}}&=&i\p_{\ove{z}}(\ln g)\tau_{1} =\frac{i}{2}\frac{e^{i\theta}}{r}\left (\nu-1\pm 2\nu\frac{r^{2\nu}}{1\mp r^{2\nu}}\right )\tau_{1}, \no\\
\Phi &=&g(i)^{n}(\tau_{2}-i\tau_{3})= 2(i)^{n}\nu \frac{r^{\nu -1}}{1\mp r^{2\nu}}(\tau_{2}-i\tau_{3}), \no\\
\Phi^{\ast} &=&-g(i)^{n}(\tau_{2}+i\tau_{3})= -2(i)^{n}\nu
\frac{r^{\nu -1}}{1\mp r^{2\nu}}(\tau_{2}+i\tau_{3})\label{Mosna4}.
\eeq

\subsubsection{Sinh-Gordon equation}

For this case we have $\kappa\neq 0$. We only will consider the case
corresponding to $n=0$, which is a real and singular solution. Now
we define \be g=\kappa \cosh(\alpha), \quad h=\kappa \sinh(\alpha).
\ee Then functions $f_{1}$ and $f_{2}$ are given in terms of
$\alpha$ , i.e.
\begin{gather}
f_{1}=-\p_{y}\alpha,\quad f_{2}=\p_{x}\alpha, \no \\
\p_{x}f_{2}-\p_{y}f_{1}=\frac{\kappa^{2}}{2} \sinh(2\alpha).
\end{gather}
The function $\alpha$ satisfies the sinh-Gordon equation \be
\nabla^{2}\alpha-\frac{\kappa^{2}}{2} \sinh(2\alpha)=0. \ee An
analytic solution of this equation is given by \cite{Dunajski12} \be
\alpha =2
\tanh^{-1}\left[\exp\left(\kappa\left(\frac{(z-z_{0})e^{-i\omega}}{2}+\frac{(\ove{z}-\ove{z}_{0})e^{i\omega}}{2}\right)\right)\right],\label{Alfa}
\ee with $z_{0}$ complex constant and $\omega$ real constant. In
this case, the Hitchin's connections are
$$
A_{z}=-i\p_{z}(\alpha)\tau_{1},  \ \ \ \ \ \ \ \ \ \
\Phi=\kappa(\cosh(\alpha)\tau_{2}-i \sinh(\alpha)\tau_{3}),
$$
\begin{equation}
A_{\ove{z}}=i\p_{\ove{z}}(\alpha)\tau_{1}, \ \ \ \ \ \ \ \ \ \
\Phi^{\ast}=-\kappa(\cosh(\alpha)\tau_{2}+i \sinh(\alpha)\tau_{3}),
\label{Alpha4}
\end{equation}
and $\alpha$ is given by (\ref{Alfa}).

\subsubsection{Painlev\'e III equation}

We now discuss a solution given by Ward in Ref. \cite{Ward2}. Ward
considered smooth SU$(2)$ solutions of the Hitchin's equations on
$\mathbb{R}^{2}$ with boundary conditions involving an integer $n$,
which is the degree of the determinant of the Higgs field $\Phi$.
Then we take the following ansatz for the fields
\begin{gather}
A_{\bz}=\f{i}{2}(\p_{\bz}\psi)\tau_{3}+\alpha\Phi, \quad A_{z}=-\f{i}{2}(\p_{z}\psi)\tau_{3}-\bar{\alpha}\Phi^{\ast}, \no\\
\Phi=\left(
\begin{array}{cc}
0 & \mu_{+}e^{\psi/2} \\
\mu_{-}e^{-\psi/2} & 0
\end{array}
\right),
\end{gather}
where the functions $\m_{+}$, $\m_{-}$, $\psi$ and $\alpha$ are
functions of the complex variables $z$ and $\bz$. The fields $A_{z},
A_{\ove{z}}, \Phi, \Phi^{\ast}$ satisfy Hitchin's equations if the
four functions fulfil the following equations
\begin{gather}
\Delta\psi=2(1+4\vert\alpha\vert^{2})(\vert\m_{+}\vert^{2}e^{\psi}-\vert\m_{-}\vert^{2}e^{-\psi}),\label{HitchinA3}\\
0=e^{-\psi/2}\p_{z}(e^{\psi}\m_{+}\alpha)+e^{\psi/2}\p_{\bz}(e^{-\psi}\bar{\m}_{-}\bar{\alpha})\label{HitchinA4},
\end{gather}
where $\Delta=4\p_{z}\p_{\bz}$. In the special case $\alpha=0$ the previous equations are locally equivalent to the sinh-Gordon equation, at the cost of more complicate global condition on the field $\psi$.

For future reasons, it is convenient to write $\Phi$ in the basis of
the Lie algebra $\mathfrak{su}(2)$, this yields \beq \Phi&=&\left(
\begin{array}{cc}
0 & \mu_{+}e^{\psi/2} \nonumber \\
\mu_{-}e^{-\psi/2} & 0
\end{array}
\right) \nonumber\\
&=&
-i(\mu_{+}e^{\psi/2}+\mu_{-}e^{-\psi/2})\f{i}{2}\left(\begin{array}{cc}
0 & 1 \nonumber\\
1 & 0
\end{array}\right)
+(\mu_{+}e^{\psi/2}-\mu_{-}e^{-\psi/2})\f{1}{2}\left(\begin{array}{cc}
0 & 1 \nonumber\\
-1 & 0
\end{array}\right) \nonumber\\
&=&f_{1}\tau_{1}+f_{2}\tau_{2}, \eeq where
\begin{equation}
f_{1}=-i(\mu_{+}e^{\psi/2}+\mu_{-}e^{-\psi/2}), \ \ \ \ \ \ \ \
f_{2}=\mu_{+}e^{\psi/2}-\mu_{-}e^{-\psi/2}.
\end{equation}

The simplest case is when the determinant of $\Phi$ is $z$, i.e.
$n=1$. In this situation the $\psi$ is only function of $r$ and, by
symmetries we have $\alpha=0$. Consequently we have the following
definitions \be \m_{+}=z,\quad \m_{-}=-1,\quad \psi=\psi(r),\quad
r=\vert z\vert. \label{valores} \ee

Thus equation (\ref{HitchinA4}) is trivial and (\ref{HitchinA3})
reduces to \be \Delta \psi=4\psi_{,z\ove{z}}=2(\vert
z\vert^{2}e^{\psi}-e^{-\psi}).\label{condi} \ee Now defining
$t=r^{3/2}$ and $h(t)=e^{-\psi/2}t^{-1/3}$, we have \be
h''-\frac{(h')^{2}}{h}+\f{h'}{t}+\f{4}{9h}-\f{4h^{3}}{9}=0, \ee
which is the Painlev\'e III equation. In this case fields take the form
$$
A_{\bz}=\f{i}{2}\psi,_{\bz}\tau_{3}, \ \ \ \ \ \ \ \ \ \ A_{z}=-\f{i}{2}\psi,_{z}\tau_{3}, \no \label{Ward1}\\
$$
\begin{equation}
\Phi=f_{1}\tau_{1}+f_{2}\tau_{2},    \ \ \ \ \ \ \ \ \ \
\Phi^{\ast}=-\bar{f}_{1}\tau_{1}-\bar{f}_{2}\tau_{2},\label{Ward4}
\end{equation}
with the functions $f_{1}$ and $f_{2}$ given by
\begin{equation}
f_{1}=-i(ze^{\f{\psi}{2}}-e^{-\f{\psi}{2}}), \ \ \ \ \ \ \ \ \ \
f_{2}=ze^{\f{\psi}{2}}+e^{-\f{\psi}{2}}. \label{f2simp}
\end{equation}

\section{SU$(\infty)$ Hitchin's equations}

In this section we obtain the SU$(\infty)$ Hitchin's equations.
Taking particular values for the hamiltonian functions, the
SU$(\infty)$ Hitchin's equations reduce to Husain-Park equation.
These particular values for the hamiltonian functions are inspired
in the reduced Hitchin system (\ref{Hitchinredu}). Solutions to the
SU$(\infty)$ Hitchin's equations are obtained from solutions of the
SU$(2)$ Hitchin's equations using WWMG-correspondence. We obtain
three different sets of hamiltonian functions and their hamiltonian
vector fields corresponding with the Liouville, sinh-Gordon and
Painlev\'e equations respectively.

Consider the case with gauge group SU$(N)$. In the large-$N$ limit,
$(N\to\infty)$, the Lie group can be written in terms of the
area-preserving diffeomorphism group of a certain surface $\Sigma$
i.e.
$$
SU(N)\to {\rm SDiff}(\Sigma),
$$
with $A_{i}\in {\rm
sdiff}(\Sigma)$. Then $A_{i}$ are hamiltonian vector fields over
$\Sigma$ and they have the following form \be
A_{i}=A_{\theta_{i}}=\theta_{i},_{p}\p_{q}-\theta_{i},_{q}\p_{p},
\label{A2} \ee where $p$ and $q$ are the local coordinates on the
two-manifold $\Sigma$, $\theta_i$ are the hamiltonian functions and
$\theta_{i},_{p}=\p_{p}\theta_{i}$,
$\theta_{i},_{q}=\p_{q}\theta_{i}$. In a more compact form, we can
use the antisymmetric matrix $\varepsilon=
\begin{pmatrix} 0 & -1 \\
1 & 0 \end{pmatrix}$ and rewrite $A_{i}$ as follows \be
A_{i}=A_{\theta_{i}}=\varepsilon^{\m\n}\theta_{i,\m}\p_{\n}.
\label{A} \ee Now the Lie bracket of two vector fields is \beq \lc
A_{i}, A_{j}\rc
&=&(\varepsilon^{\m\n}\theta_{i,\m}\varepsilon^{\sigma\rho}\theta_{j,\sigma\n}-
\varepsilon^{\m\n}\theta_{j,\m}\varepsilon^{\sigma\rho}\theta_{i,\sigma\n})\p_{\rho}\no\\
&=& \varepsilon^{\sigma\rho}\{\theta_{i},\theta_{j}\},_{\sigma}\p_{\rho}\no\\
&=&A_{\{\theta_{i},\theta_{j}\}}, \eeq where
$\{\theta_{i},\theta_{j}\}=\theta_{i,p}\theta_{j,q}-\theta_{i,q}\theta_{j,p}$
is the Poisson bracket over $\mathbb{R}^2$.

Hitchin's equations (\ref{HitchinA1}) and (\ref{HitchinA2}), and
also $\p_{z}\Phi^{\ast}+[A_{z},\Phi^{\ast}]=0$, with the gauge group
SDiff$(\Sigma)$ have the form
\begin{gather}
H,_{\bz}+\{ H_{\bz},H\}=0, \quad H_{\ast},_{z}+\{ H_{z},H_{\ast}\}=0,\label{Hitchingrav4}\\
H_{\bz,z}-H_{z,\bz}+\{ H_{z},H_{\bz}\}+\f{1}{4}\{ H,
H_{\ast}\}=0,\label{Hitchingrav3}
\end{gather}
where $H_{z}, H_{\bz}, H, H_{\ast}$ are the hamiltonian functions
depending on the coordinates $(z,\ove{z},p,q)$ associated to $A_{z},
A_{\bz}, \Phi, \Phi^{\ast}$ respectively. Right hand side of Eqs.
(\ref{Hitchingrav4}) and (\ref{Hitchingrav3}) are actually equal to
three corresponding arbitrary functions $f_{i}$ $(i=1,2,3)$,
depending only on the coordinates $(z,\ove{z})$. In the rest of this
paper we only consider the case $f_{i}=0$.

\subsection{Husain-Park equation}

If we take the particular case when $H=2H_{z}$ and
$H_{\ast}=2H_{\bz}$, then we have the simplified system
\begin{equation}
H_{z},_{\bz}-\{H_{z},H_{\bz}\}=0, \quad H_{z},_{\bz}+H_{\bz},_{z}=0.
\label{Hitchingravred}
\end{equation}
Furthermore we can use the integrability condition and take
$H_{z}=\Lambda,_{z}$ and $H_{\bz}=-\Lambda,_{\bz}$ with
$\Lambda=\Lambda(z,\bz,p,q)$. In this case we obtain a second order
equation
\begin{equation}
\Lambda,_{z\bz}+\{\Lambda,_{z}, \Lambda,_{\bz}\}=0,
\end{equation}
which is a complex version of the Husain-Park equation
\cite{Compean}. Indeed, if we take $z=x+iy$, $\Lambda\to i\Lambda$
and a rescaling in the coordinates $p$ and $q$ we arrive to
\begin{equation}
\Lambda,_{yy}+\Lambda,_{xx}+\Lambda,_{yp}\Lambda,_{xq}-\Lambda,_{yq}\Lambda,_{xp}=0.
\label{HPE}
\end{equation}
Notice that equations (\ref{Hitchingravred}) are equations
(\ref{Hitchinredu}) when the gauge group is SDiff$(\Sigma).$
Consequently for a particular value of the spectral parameter, the
Hitchin's equations with gauge group SDiff$(\Sigma)$ describe the
complex version of a Husain-Park equation.

\subsection{Weyl-Wigner-Moyal-Groenewold correspondence}

In this section we survey the Weyl-Wigner-Moyal-Groenewold (WWMG)
correspondence in order to use this to find solutions of the
SU$(\infty)$ Hitchin's equations. Following Ref. \cite{Compean}, we
use Moyal deformation to find solutions of (\ref{Hitchingrav3}) and
(\ref{Hitchingrav4}). First of all we associate to each Lie
algebra-valued element $K$ a corresponding unitary operator
$\widehat{K}$ acting on a Hilbert space. Then, using the WWMG
correspondence, we can associate to $\widehat{K}$ a function
$\widetilde{K}=\widetilde{K}(z,\ove{z},p,q)$ by \be
\widetilde{K}=\int_{-\infty}^{\infty}\bigg\langle
p-\f{\zeta}{2}\vert \widehat{K} \vert p+\f{\zeta}{2} \bigg\rangle
\exp \left(\f{iq\zeta}{\hbar}\right)d\zeta.\label{integral} \ee For
a commutator of two operators, we have \be \{\widetilde{K}_{1},
\widetilde{K}_{2}\}_{M} := \int_{-\infty}^{\infty}\bigg\langle
p-\f{\zeta}{2}\bigg\vert \lc \widehat{K}_{1}, \widehat{K}_{2}\rc
\bigg\vert p+\f{\zeta}{2} \bigg\rangle \exp
\left(\f{iq\zeta}{\hbar}\right)d\zeta. \ee Right hand side defines
the Moyal bracket which can be expressed in a more practical way \be
\{\widetilde{K}_{1}, \widetilde{K}_{2}\}_{M} = \f{2}{\hbar}
\widetilde{K}_{1}\sin\left(\f{\hbar}{2}
\buildrel{\longleftrightarrow} \over{P}\right)\widetilde{K}_{2}, \ee
where \be \buildrel{\longleftrightarrow} \over {P}=
\overleftarrow{\f{\p}{\p p}} \overrightarrow{\f{\p}{\p
q}}-\overleftarrow{\f{\p}{\p q}}\overrightarrow{\f{\p}{\p p}}. \ee

Using the previous procedure we can associate functions
$\widetilde{H}_{z}$, $\widetilde{H}_{\ove{z}}$, $\widetilde{H}$,
$\widetilde{H}_{\ast}$ to fields $A_{z}, A_{\bz}, \Phi, \Phi^{\ast}$
respectively and the Hitchin's equations take the form
\begin{gather}\widetilde{H}_{\ove{z},z}-\widetilde{H}_{z,\ove{z}}+\{\widetilde{H}_{z},\widetilde{H}_{\ove{z}}\}_{M}+\f{1}{4}\{\widetilde{H}, \widetilde{H}_{\ast}\}_{M}=0,\\
\widetilde{H},_{\ove{z}}+\{ \widetilde{H}_{\ove{z}}, \widetilde{H}\}_{M}=0.
\end{gather}

Taking the limit $\hbar\to 0$, the Moyal bracket reduces to the
Poisson bracket \be \lim_{\hbar\to
0}\{\cdot,\cdot\}_{M}=\{\cdot,\cdot\}. \ee We have assumed that the
function $\widetilde{K}$ is an analytic function of $\hbar$, i.e.
\beq \widetilde{K}=\sum_{n=0}^{\infty}\hbar^{n}K_{n},\ \ \ \ \ \ \
K_{n}=K_{n}(z,\ove{z},p,q). \eeq Then the functions obtained from
the Moyal deformation provides the hamiltonian functions which are
solutions of the SU$(\infty)$ Hitchin's equations
(\ref{Hitchingrav4}) and (\ref{Hitchingrav3}), in the limit
$\lim_{\hbar\to 0}\widetilde{H}=H_{0}=H$, etc.

\subsection{Solutions}

Using the WWMG-correspondence, we can obtain solutions to the
SU$(\infty)$ Hitchin's equations, (\ref{Hitchingrav4}) and (\ref{Hitchingrav3}), from solutions to the SU$(2)$ Hitchin's equations considered previously as: the Liouville, sinh-Gordon and Painlev\'e III equations. In these cases we have
\begin{equation}
A(z,\bz)=A_{i}(z,\bz)\tau_{i},
\end{equation}
for the Yang-Mills field. In the previous equation, summation over
the repeated indices should be understood. Now we consider that
$\tau_{i}$, a basis for the $\mathfrak{su}(2)$, can be substituted by the corresponding basis of self-dual operators over the Hilbert space $\widehat{\tau}_{i}$. The explicit form of the $\widehat{\tau}_{i}$ are
\begin{gather}
\widehat{\tau}_{1}=i\beta \widehat{q}+\frac{1}{2\hbar}(\widehat{q}^{2}-1)\widehat{p}, \quad \widehat{\tau}_{2}=-\beta \widehat{q}+\frac{i}{2\hbar}(\widehat{q}^{2}+1)\widehat{p}, \no \\
\widehat{\tau}_{3}=-i\beta
\widehat{1}-\frac{1}{\hbar}\widehat{q}\widehat{p}.
\end{gather}

Using the WWMG-correspondence (\ref{integral}), these operators
$\widehat{\tau}_{i}$ is mapped to functions $X_{i}(p,q)$ in the
coordinates $p,q$ of $\Sigma$. Thus, the field $A(z,\bz)$ is now a
function of all the coordinates $A(z,\bz,p,q)=i\hbar
A_{i}(z,\bz)X_{i}(p,q)$ given by
\begin{equation}
\frac{i}{2}A_{1}
p(q^{2}-1)-\frac{1}{2}A_{2}p(q^{2}+1)-iA_{3}pq+\hbar\left(\beta+
\frac{1}{2}\right)(-A_{1}q-iA_{2}q+A_{3}).
\end{equation}
In a limit when $\hbar \to 0$, the previous solution leads to a
solution for the SU$(\infty)$ Hitchin's equations
(\ref{Hitchingrav4}) and (\ref{Hitchingrav3}).

Now we use the solutions of the SU$(2)$ Hitchin's equation,
expressed in terms of Liouville, sinh-Gordon and Painlev\'e III
equations, to find three different sets of hamiltonian functions
which are solutions of SU$(\infty)$ Hitchin's equations. We also
build three different sets of hamiltonian vector fields and in
section 5 we use them to build half-flat metrics.

\subsubsection{Liouville equation}

From solutions (\ref{Mosna4}) to Hitchin's equations, the
corresponding hamiltonian function can be obtained by using the
WWMG-correspondence. They are given by
$$
H_{z}=-\frac{1}{4}p(q^{2}-1)e^{-i\theta}\p_{r}(\ln g), \ \ \ \ \ \ \ \ \ \ H=\frac{(i)^{n}gp}{2}(q+1)^{2}, \no \label{Mosna5}\\
$$
\begin{equation}
H_{\ove{z}}=\frac{1}{4}p(q^{2}-1)e^{i\theta}\p_{r}(\ln g), \ \ \ \ \
\ \ \ \ \ H_{\ast}=-\frac{(i)^{n}gp}{2}(q-1)^{2},\label{Mosna8}
\end{equation}
which are solutions of the Hitchin's equations (\ref{Hitchingrav3})
and (\ref{Hitchingrav4}). The corresponding hamiltonian vector
fields are given by \beq
A_{z}&=&H_{z,p}\p_{q}-H_{z,q}\p_{p}=-\frac{1}{4}(q^{2}-1)e^{-i\theta}\p_{r}(\ln g)\p_{q}+\frac{1}{2}pqe^{-i\theta}\p_{r}(\ln g) \p_{p}, \no \\
A_{\ove{z}}&=&H_{\ove{z},p}\p_{q}-H_{\ove{z},q}\p_{p}=\frac{1}{4}(q^{2}-1)e^{i\theta}\p_{r}(\ln g)\p_{q}-\frac{1}{2}pqe^{i\theta}\p_{r}(\ln g)\p_{p}, \no\\
\Phi&=&H_{,p}\p_{q}-H_{,q}\p_{p}=\frac{(i)^{n}g}{2}(q+1)^{2}\p_{q}-(i)^{n}gp(q+1)\p_{p}, \no\\
\Phi^{\ast}&=&H_{\ast,p}\p_{q}-H_{\ast,q}\p_{p}=-\frac{(i)^{n}g}{2}(q-1)^{2}\p_{q}+(i)^{n}gp(q-1)\p_{p}.
\eeq

\subsubsection{Sinh-Gordon equation}

In this case, the solutions are given in Eqs. (\ref{Alpha4}).
Consequently the hamiltonian functions obtained from the sinh-Gordon
equation are
$$
H_{z}=-\frac{1}{2}p(q^{2}-1)\alpha,_{z}, \ \ \ \ \ \ \ \ \ \ H=-\kappa p\left[\frac{1}{2}(q^{2}+1)\cosh(\alpha)+q \sinh(\alpha)\right], \no \label{Alpha5}\\
$$
\begin{equation}
H_{\ove{z}}=\frac{1}{2}p(q^{2}-1)\alpha,_{\ove{z}}, \ \ \ \ \ \ \ \
\ \ H_{\ast}=-\kappa p\left[-\frac{1}{2}(q^{2}+1)\cosh(\alpha)+q
\sinh(\alpha)\right].\label{Alpha8}
\end{equation}

The hamiltonian vectors fields are given by \beq
A_{z}&=&-\frac{1}{2}(p^{2}-1)\alpha,_{z}\p_{q}+pq\alpha,_{z}\p_{p}, \no\\
A_{\ove{z}}&=&\frac{1}{2}(p^{2}-1)\alpha,_{\ove{z}}\p_{q}-pq\alpha,_{\ove{z}}\p_{p}, \no\\
\Phi &=& -\kappa\big[\frac{1}{2}(q^{2}+1)\cosh(\alpha)+q \sinh(\alpha)\big]\p_{q}+\kappa p\big[q \cosh(\alpha)+ \sinh(\alpha)\big]\p_{p},\no\\
\Phi^{\ast} &=& -\kappa\big[-\frac{1}{2}(q^{2}+1) \cosh(\alpha)+q \sinh(\alpha)\big]\p_{q}+\kappa p\big[-q \cosh(\alpha)+ \sinh(\alpha)\big]\p_{p}.\no\\
\eeq

\subsubsection{Painlev\'e III equation}

In the case of the Painlev\'e III equation, the solutions are given
by Eqs. (\ref{Ward4}) and we find that the hamiltonian functions are
given by
\begin{gather}
H_{z}=-\f{1}{2}pq\psi,_{z}, \quad H_{\bz}=\f{1}{2}pq\psi,_{\bz}, \no \label{Sol1}\\
H=\f{i}{2}p(q^{2}-1)f_{1}-\f{1}{2}p(q^{2}+1)f_{2}=-e^{-\f{\psi}{2}}p(q^{2}+e^{\psi}z), \no \label{Sol3}\\
H_{\ast}=-\f{i}{2}p(q^{2}-1)\bar{f}_{1}+\f{1}{2}p(q^{2}+1)\bar{f}_{2}=e^{-\f{\psi}{2}}p(1+e^{\psi}q^{2}\ove{z}),\label{Sol4}
\end{gather}
where the functions $f_{1}$ and $f_{2}$ are given by (\ref{f2simp}).
These hamiltonian functions are solutions of Hitchin's equations
(\ref{Hitchingrav3}) and (\ref{Hitchingrav4}), where the condition
(\ref{condi}) is satisfied. The hamiltonian vector fields are \beq
A_{z}&=&H_{z,p}\p_{q}-H_{z,q}\p_{p}=-\f{1}{2}q\psi,_{z}\p_{q}+\f{1}{2}p\psi,_{z}\p_{p}, \no\\
A_{\ove{z}}&=&H_{\ove{z},p}\p_{q}-H_{\ove{z},q}\p_{p}=\f{1}{2}q\psi,_{\ove{z}}\p_{q}-\f{1}{2}p\psi,_{\ove{z}}\p_{p}, \no \\
\Phi&=&H_{,p}\p_{q}-H_{,q}\p_{p}=-e^{-\f{\psi}{2}}(q^{2}+e^{\psi}z)\p_{q}+2e^{-\f{\psi}{2}}pq\p_{p}, \no\\
\Phi^{\ast}&=&H_{\ast,p}\p_{q}-H_{\ast,q}\p_{p}=-e^{-\f{\psi}{2}}(1+e^{\psi}q^{2}\ove{z})\p_{q}+2e^{\f{\psi}{2}}pq
\ove{z}\p_{p}. \eeq

\section{Conformal compactification case of SU$(\infty)$ Hitchin's equations}

In this section we represent solutions of the SU$(2)$ Hitchin's
equations as hamiltonian vector fields. To do this we need reexpress
$\tau_{i}$, the basis of Lie algebra $\mathfrak{su}(2)$, as
hamiltonian vector fields $X_{H_{i}}$ on $\mathbb{CP}^{1}$. From
this we build three different sets of hamiltonian vector fields
depending of three different equations: Liouville, sinh-Gordon and
Painlev\'e III. In the following section we use this information to
build half-flat metrics.

As we mentioned before, Mason conjectured that SL$(N,\mathbb{C})$ is
a subgroup of SL$(\infty)$ only for $N=2$. In Ref.
\cite{Dunajski98}, it was found a way to reproduce self-dual gravity
from the linear action of SU$(2)$ on $\mathbb{CP}^1$. In this case,
the covariant derivative can be expressed in terms of the
hamiltonian vector fields and it is given by \be
D_{i}=\p_{i}+A_{i}=\p_{i}+A_{i}^{k}X_{H_{k}}, \ee where $X_{H_{i}}$
are the hamiltonian vector fields of $\mathfrak{su}(2)$ over
$\mathbb{CP}^{1}$ and $H_{i}$ are hamiltonian functions. Moreover
$\mathbb{CP}^{1}$ has a symplectic form given by \be
\Omega_{\mathbb{CP}^{1}}=\Omega dp\wedge dq,  \ee with
$\Omega=\f{i}{(1+pq)^{2}}$ and the hamiltonian functions are given
by \be H_{1}=-\f{p+q}{1+pq},\quad H_{2}=-i\f{p-q}{1+pq},\quad
H_{3}=\f{2}{1+pq}, \ee where $p$ and $q$ are local coordinates over
$\mathbb{CP}^{1}$. The hamiltonian vector fields can be written as
\be
X_{H_{i}}=X_{H_{i}}^{a}\p_{a}=\f{1}{\Omega}\varepsilon^{ab}H_{i,a}\p_{b}=\f{1}{\Omega}\left(H_{i,p}\p_{q}-H_{i,q}\p_{p}\right).
\ee Explicitly we have
\begin{gather}
X_{H_{1}}=-i(q^{2}-1)\p_{q} +i(p^{2}-1)\p_{p},\quad X_{H_{2}}=-(q^{2}+1)\p_{q}-(p^{2}+1)\p_{p}, \no \label{Hamil1}\\
X_{H_{3}}=2i(q\p_{q}-p\p_{p})\label{Hamil3}.
\end{gather}

Hitchin's equations which fulfil the previous hamiltonian vector
fields are given by
\begin{gather}
\p_{z}A_{\bz}-\p_{\bz}A_{z}+\frac{1}{2}\lc A_{z},A_{\bz}\rc+\f{1}{8}\lc \Phi, \Phi^{\ast}\rc=0,\label{HitchinFac1}\\
\p_{\bz}\Phi+\f{1}{2}\lc A_{\bz},\Phi\rc=0\label{HitchinFac2},
\end{gather}
where the bracket is the Lie bracket of two vector fields.

\subsection{Liouville equation}

In the case of the Liouville equation, the hamiltonian vector fields
have the form \beq
A_{z}&=&-\frac{1}{2}(q^{2}-1)e^{-i\theta}\p_{r}(\ln g)\p_{q}+\frac{1}{2}(p^{2}-1)e^{-i\theta}\p_{r}(\ln g)\p_{p}, \no\\
A_{\ove{z}}&=&\frac{1}{2}(q^{2}-1)e^{i\theta}\p_{r}(\ln g)\p_{q}-\frac{1}{2}(p^{2}-1)e^{i\theta}\p_{r}(\ln g)\p_{p}, \no\\
\Phi &=&-g(i)^{n}(q-1)^{2}\p_{q}-g(i)^{n}(p+1)^{2}\p_{p}, \no\\
\Phi^{\ast} &=&g(i)^{n}(q+1)^{2}\p_{q}+g(i)^{n}(p-1)^{2}\p_{p}. \eeq
They are solutions of equations (\ref{HitchinFac1}) and
(\ref{HitchinFac2}) if $g$ has the form described in Eq. (\ref{g}).

\subsection{Sinh-Gordon equation}

For the case of the sinh-Gordon equation, the hamiltonian vector
fields are
\begin{gather}
A_{z}=(q^{2}-1)\alpha,_{z}\p_{q}-(p^{2}-1)\alpha,_{z}\p_{p}, \no\\
A_{\ove{z}}=-(q^{2}-1)\alpha,_{\ove{z}}\p_{q}+(p^{2}-1)\alpha,_{\ove{z}}\p_{p}, \no\\
\Phi=\kappa[2q \sinh(\alpha)-(q^{2}+1) \cosh(\alpha)]\p_{q}-\kappa[2p \sinh(\alpha)+(p^{2}+1) \cosh(\alpha)]\p_{p}, \no\\
\Phi^{\ast}=\kappa[2q \sinh(\alpha)+(q^{2}+1)
\cosh(\alpha)]\p_{q}+\kappa[-2p \sinh(\alpha)+(p^{2}+1)
\cosh(\alpha)]\p_{p},
\end{gather}
where $\alpha$ has the form given in Eq. (\ref{Alfa}).

\subsection{Painlev\'e III equation}

In the case of the Painlev\'e III equation, we obtain the following
hamiltonian vector fields
\beq
A_{z}&=&\psi_{,z}q\p_{q}-\psi_{,z}p\p_{p}, \no\\
A_{\ove{z}}&=&-\psi_{,\ove{z}}q\p_{q}+\psi_{,\ove{z}}p\p_{p}, \no\\
\Phi&=&\big[-if_{1}(q^{2}-1)-f_{2}(q^{2}+1)\big]\p_{q}+ \big[if_{1}(p^{2}-1)-f_{2}(p^{2}+1)\big]\p_{p}\no\\
&=&-2e^{-\f{\psi}{2}}(1+e^{\psi}q^{2}z)\p_{q}+2e^{-\f{\psi}{2}}(p^{2}+e^{\psi}z)\p_{p}, \no\\
\Phi^{\ast}&=&\big[i\ove{f_{1}}(q^{2}-1)+\ove{f_{2}}(q^{2}+1)\big]\p_{q}+ \big[-i\ove{f_{1}} (p^{2}-1)+\ove{f_{2}}(p^{2}+1)\big]\p_{p}\no\\
&=&2e^{-\f{\psi}{2}}(q^{2}+e^{\psi}\ove{z})\p_{q}-2e^{-\f{\psi}{2}}(1+e^{\psi}p^{2}z)\p_{q},
\eeq with $f_{1}$ and $f_{2}$ given by Eqs. (\ref{f2simp})
respectively, and $\psi$ is a solution of (\ref{condi}).

\section{Self-dual gravity}

In this section we finally build half-flat metrics on
$\mathcal{M}_{1}\times\mathcal{M}_{2}$, where $\mathcal{M}_{i}$ are
two-dimensional manifolds. We make this using the Mason and Newman
formulation of SDG which connect the Lax pair formalisms of SDYM and
SDG. The half-flat metrics are coming from hamiltonian vector fields
which solve the SU$(\infty)$ Hitchin's equations. In the case of the
infinite dimensional gauge group we first give the most general
half-flat metric and then show that this metric is reduced to
Husain's metric with a specific choose of the hamiltonian functions.
Later we build six different half-flat metrics using the Liouville,
sinh-Gordon and Painlev\'e equations. In each case also the dual
frame is given.

In Ref. \cite{Mason89}, Mason and Newman shown that given four
independent vectors fields $V_{a}=(W,\widetilde{W}, Z,
\widetilde{Z})$ over a complex four-manifold $M$, and given a
non-zero four-form $\nu$ (volume form) which satisfies \be
[L,M]=0,\quad \mathcal{L}_{L}\nu=-\mathcal{L}_{M}\nu=0, \label{Lax}
\ee where ${\cal L}$ is the Lie derivative,
$L=Z-\lambda\widetilde{W}$ and $M=W-\lambda \widetilde{Z}$. Then
$f^{-1}V_{a}$ determines a null-tetrad for a half-flat metric (i.e.
with vanishing Ricci tensor and self-dual Weyl tensor), where
$f^{2}=\nu(W,\widetilde{W},Z,\widetilde{Z})$. Moreover, every
half-flat metric arises in this way.

Consequently Eq. (\ref{Lax}) impose the following conditions over
the vector fields $V_{a}$ \be [W,Z]=0,\quad
[\widetilde{W},\widetilde{Z}]=0, \quad
[W,\widetilde{W}]+[\widetilde{Z},Z]=0.\label{Lax2} \ee If we make
the identifications \be Z=D_{z},\quad
\widetilde{Z}=D_{\ove{z}},\quad W=\frac{1}{2}\Phi^{\ast},\quad
\widetilde{W}=\frac{1}{2}\Phi, \ee then Eqs. (\ref{Lax2}) expresses
the Hitchin's equations (\ref{HitchinA1}) and (\ref{HitchinA2}).
This shown that the vector fields
$f^{-1}(D_{z},D_{\ove{z}},\frac{1}{2}\Phi,\frac{1}{2}\Phi^{\ast})$
form a null-tetrad for a half-flat metric.

In this case the metric is expressed in the dual frame $e_{V_{a}}$
\be g=f^{2}(e_{Z}\odot e_{\widetilde{Z}}-e_{W}\odot
e_{\widetilde{W}}),\label{Metrica} \ee where $\odot$ stands for the
symmetrized tensor product and
$$
e_{W}=f^{-2}\nu(-,\widetilde{W},Z,\widetilde{Z}), \ \ \ \ \ \ \ \ \ \ \ \ \ e_{\widetilde{W}}=f^{-2}\nu(W,-,Z,\widetilde{Z})\label{Marco1},\\
$$
\begin{equation}
e_{Z}=f^{-2}\nu(W,\widetilde{W} ,-,\widetilde{Z}), \ \ \ \ \ \ \ \ \
\ \ \ \ e_{\widetilde{Z}}=f^{-2}\nu(W,\widetilde{W}
,Z,-)\label{Marco4}.
\end{equation}

In the case of the Hitchin's vector fields, the dual frame
(\ref{Marco4}) and the function $f$, are given by
\begin{gather}
e_{\Phi^{\ast}}=f^{-2}\nu\left (-,\frac{1}{2}\Phi, D_{z}, D_{\ove{z}}\right ), \quad e_{\Phi}=f^{-2}\nu\left (\frac{1}{2}\Phi^{\ast},-,  D_{z}, D_{\ove{z}}\right ), \no \label{Marco5}\\
e_{Z}=f^{-2}\nu\left (\frac{1}{2}\Phi^{\ast},\frac{1}{2}\Phi,-, D_{\ove{z}}\right ), \quad e_{\ove{Z}}=f^{-2}\nu\left (\frac{1}{2}\Phi^{\ast}, \frac{1}{2}\Phi, D_{z},-\right ), \no \label{Marco8}\\
f^{2}=\nu\left
(\frac{1}{2}\Phi^{\ast},\frac{1}{2}\Phi,D_{z},D_{\ove{z}}\right ).
\label{Marco8}
\end{gather}

\subsection{Infinite gauge group}

Now we construct half-flat metrics coming from solutions of the
SU$(\infty)$ Hitchin's equations obtained by Moyal deformation. The
four-form is given by $\nu=dz\wedge d\ove{z}\wedge dp\wedge dq$ and
the hamiltonian vector fields are
$$
D_{z}=\p_{z}+\varepsilon^{\mu\nu}H_{z,\mu}\p_{\nu}, \ \ \ \ \ \ \ \ \ \ \ \ \ \ \ \Phi=\varepsilon^{\mu\nu}H,_{\mu}\p_{\nu},\\
$$
\begin{equation}
D_{\ove{z}}=\p_{\ove{z}}+\varepsilon^{\mu\nu}H_{\ove{z},\mu}\p_{\nu},
\ \ \ \ \ \ \ \ \ \ \ \ \ \ \
\Phi^{\ast}=\varepsilon^{\mu\nu}H_{\ast,\mu}\p_{\nu},
\end{equation}
where $H_{z},H_{\ove{z}},H,H_{\ast}$ are solutions of equations
(\ref{Hitchingrav3}) and (\ref{Hitchingrav4}). In this case the dual
frame (\ref{Marco8}) and the function $f$ are
\begin{gather}
e_{\Phi^{\ast}}=\frac{f^{-2}}{2}\big[H,_{q}Dq+H,_{p}Dp\big], \quad e_{z}=dz, \no\\
e_{\Phi}=-\frac{f^{-2}}{2}\big[H_{\ast, q}Dq+H_{\ast ,p}Dp\big], \quad e_{\ove{z}}=d\ove{z}, \no \\
f^{2}=\frac{1}{4}\{H_{\ast},H\},
\end{gather}
where $Dp$ and $Dq$ are defined as follows \be
Dp:=dp+H_{\mu,q}dx^{\mu}, \quad Dq:=dq-H_{\mu,p}dx^{\mu}, \ee with
$H_{\mu}dx^{\mu}=H_{z}dz+H_{\ove{z}}d\ove{z}$. Moreover, the metric
is given by
\begin{gather}
ds^{2}=\frac{1}{\{H_{\ast},H\}} \bigg\{ \{H,H_{z}\}\{H_{\ast},H_{z}\}dz^{2}+\{H,H_{\ove{z}}\}\{H_{\ast},H_{\ove{z}}\}d\ove{z}^{2}+H,_{p}H_{\ast},_{p}dp^{2}+H,_{q}H_{\ast},_{q}dq^{2}\no\\
+\left(\{H,H_{z}\}\{H_{\ast},H_{\ove{z}}\}+\{H_{\ast},H_{z}\}\{H,H_{\ove{z}}\}+\frac{\{H_{\ast},H\}^{2}}{4}\right)dzd\ove{z}+\left(H,_{p}H_{\ast},_{q}+H,_{q}H_{\ast},_{p}\right)dpdq\no\\
+\left(H,_{p}\{H_{\ast},H_{z}\}+H_{\ast},_{p}\{H,H_{z}\}\right)dzdp+\left(H,_{q}\{H_{\ast},H_{z}\}+H_{\ast},_{q}\{H,H_{z}\}\right)dzdq\no\\
+\left(H,_{p}\{H_{\ast},H_{\ove{z}}\}+H_{\ast},_{p}\{H,H_{\ove{z}}\}\right)d\ove{z}dp+\left(H,_{q}\{H_{\ast},H_{\ove{z}}\}+H_{\ast},_{q}\{H,H_{\ove{z}}\}\right)d\ove{z}dq
\bigg\}. \label{generalHP}
\end{gather}

\subsubsection{Husain-Park metric}

For a particular value of the spectral parameter, $\lambda=1$, we
have $H=2H_{z}$, $H_{\ast}=2H_{\bz}$ and $H_{z}=\Lambda,_{z}$,
$H_{\bz}=-\Lambda,_{\bz}$, then the previous metric reduces to
\begin{multline}
ds^{2}=\left(\Lambda,_{zp}dp+\Lambda,_{zq}dq\right)dz-\left(\Lambda,_{\bz p}dp+\Lambda,_{\bz q}dq\right)d\bz\\
+\frac{1}{\{\Lambda,_{z},\Lambda,_{\bz}\}}\bigg[\Lambda,_{zp}\Lambda,_{\bz
p}dp^{2}+(\Lambda,_{zp}\Lambda,_{\bz q}+\Lambda,_{zq}\Lambda,_{\bz
p})dpdq+\Lambda,_{zq}\Lambda,_{\bz q}dq^{2}\bigg].
\end{multline}
Taking $z=x+iy$, $\Lambda\to i\Lambda$ and rescaling the coordinates
we recover the Husain metric\footnote{There is an overall factor $i$.}
\begin{multline}
ds^{2}=\left(\Lambda,_{xp}dp+\Lambda,_{xq}dq\right)dx+\left(\Lambda,_{yp}dp+\Lambda,_{yq}dq\right)dy\\
+\frac{1}{\{\Lambda,_{x},\Lambda,_{y}\}}\bigg[(\Lambda,_{xp}dp+\Lambda,_{xq}dq)^{2}+(\Lambda,_{yp}dp+\Lambda,_{yq}dq)^{2}\bigg].
\label{specialHP}
\end{multline}

\subsubsection{Liouville equation}

If we take $H_{z},H_{\ove{z}},H,H_{\ast}$ as given in Eqs.
(\ref{Mosna8}), then we obtain
\begin{gather}
Dp=dp-irpq\p_{r}(\ln g)d\theta, \quad Dq=dq+\frac{i(q^{2}-1)}{2}r\p_{r}(\ln g)d\theta, \no\\
f^{2}=\pm\frac{p(q^{2}-1)g^{2}}{4},
\end{gather}
where $g$ is given by (\ref{g}) and the $+$ sign corresponds to $n=0$, while the $-$ sign
corresponds to $n=1$. In this case $e_{\Phi}$ and $e_{\Phi^{\ast}}$
have the form \beq
e_{\Phi^{\ast}}&=&\frac{1}{i^{n}p(q-1)g}\big[2pDq+(q+1)Dp\big], \no\\
e_{\Phi}&=&\frac{1}{i^{n}p(q+1)g}\big[2pDq +(q-1)Dp\big]. \eeq

Then metric is written as
\begin{multline}
ds^{2}=\frac{p(q^{2}-1)}{4}\bigg[\pm g^{2}dr^{2}+r^{2}(\pm g^{2}-(\p_{r}(\ln g))^{2})d\theta^{2}\bigg]-pdq^{2}-\frac{q^{2}-1}{4p}dp^{2}\\
+ipr\p_{r}(\ln g)dqd\theta -qdpdq.
\end{multline}

If we take the top sign, $n=0$, we obtain a singular metric
corresponding to a real solution of the SDYM equations in two
dimensions. On the other hand, the bottom sign, $n=1$, gives rise to
a regular metric related to complex solutions of SDYM equations in
two dimensions.

As we mentioned previously, if we take the value $\nu=1$ this is
related with the abelian magnetic monopole, $M=\pm\frac{2r^{2}}{1\mp
r^{2}}d\theta$. Thus we have \be ds^{2}=\pm\frac{p(q^{2}-1)}{(1\mp
r^{2})^{2}}dr^{2}+\frac{p(q^{2}-1)}{2}Md\theta+ipMdq-pdq^{2}-\frac{q^{2}-1}{4p}dp^{2}-qdpdq.
\label{LMOne}\ee

\subsubsection{Sinh-Gordon equation}

If we define $\beta(z,\ove{z})$ in the following form \be
\beta=\kappa\left(\frac{(z-z_{0})e^{-i\omega}}{2}+\frac{(\ove{z}-\ove{z}_{0})e^{i\omega}}{2}\right),
\ee then using the hamiltonian functions (\ref{Alpha8}) we obtain
\begin{gather}
Dp=dp+\frac{1}{2}\operatorname{csch}(\beta)pq\kappa(e^{-i\omega}dz-e^{i\omega}d\ove{z}), \no\\
Dq=dq-\frac{1}{4}\operatorname{csch}(\beta)(q^{2}-1)\kappa(e^{-i\omega}dz-e^{i\omega}d\ove{z}), \no\\
f^{2}=\frac{\kappa^{2}p(q^{2}-1)}{4}\operatorname{csch}(\beta)\operatorname{coth}(\beta).
\end{gather}

The components of the dual frame $e_{\Phi^{\ast}}$ and $e_{\Phi}$
have the form \beq
e_{\Phi^{\ast}}&=&\frac{\sinh(\beta)}{\kappa p (q^{2}-1)}\bigg[2p(q+\operatorname{sech}(\beta))Dq+(1+q^{2}+2q \operatorname{sech}(\beta))Dp\bigg], \no\\
e_{\Phi}&=&\frac{\sinh(\beta)}{\kappa p (q^{2}-1)}\bigg[2p(q-
\operatorname{sech}(\beta))Dq+(1+q^{2}-2q
\operatorname{sech}(\beta))Dp\bigg].
\eeq

In the simplest case, we take $\omega=0$ and $z_{0}=0$, then the
function $\beta$ is only function of $x$, i.e.  $\beta=\kappa x$.
The metric obtained in this case is given by
\begin{multline}
ds^{2}=\frac{p(q^{2}-1)}{4}\kappa^{2} \coth(\kappa x)\operatorname{csch}(\kappa x)dx^{2}+\frac{p(q^{2}+1)}{4(q^{2}-1)}\kappa^{2} \operatorname{sech}(\kappa x)dy^{2} \\
-\frac{iq(q^{2}+1)}{q^{2}-1}\kappa \tanh(\kappa x)dpdy-\frac{2ip\kappa \big[-1+q^{2} \cosh(2\kappa x)\big] \cosh(2\kappa x)}{q^{2}-1}dqdy  \\
+\frac{4q^{2} \operatorname{sech}(\kappa x)-(q^{2}+1)^{2} \cosh(\kappa x)}{4p(q^{2}-1)}dp^{2}+\frac{2q \operatorname{sech}(\kappa x)-q(q^{2}+1) \cosh(\kappa x)}{q^{2}-1}dpdq \\
+\frac{p\big[1-q^{2}\cosh^{2}(\kappa
x)\big]\operatorname{sech}(\kappa x)}{q^{2}-1}dq^{2}. \label{SGMOne}
\end{multline}

\subsubsection{Painlev\'e III equation}

Now, we take the values of the hamiltonian functions given in
(\ref{Sol4}), then we have
\begin{gather}
Dp=dp-\frac{i}{2}pr\psi,_{r}d\theta, \quad Dq=dq+\frac{i}{2}qr\psi,_{r}d\theta, \no\\
f^{2}=\frac{1}{2}pq\left(e^{\psi}r^{2}-e^{-\psi}\right).
\end{gather}

Then, $e_{\Phi}$ and $e_{\Phi^{\ast}}$ are given by \beq
e_{\Phi^{\ast}}&=&-\frac{e^{-\frac{\psi}{2}}}{pq\left(e^{\psi}r^{2}-e^{-\psi}\right)}\bigg[2pqDq+(q^{2}+e^{\psi}re^{i\theta})Dp\bigg], \no\\
e_{\Phi}&=&-\frac{e^{\frac{\psi}{2}}}{pq\left(e^{\psi}r^{2}-e^{-\psi}\right)}\bigg[2pqre^{-i\theta}Dq+(e^{-\psi}+q^{2}re^{-i\theta})Dp\bigg].
\eeq Gathering all that information we can write down the metric as
\begin{multline}
ds^{2}=
\frac{1}{8}pq\left(e^{\psi}r^{2}-e^{-\psi}\right)(dr^{2}+r^{2}d\theta^{2})-\frac{2}{pq\left(e^{\psi}r^{2}-e^{-\psi}\right)}\bigg[e^{-\psi}\left(1+q^{2}re^{\psi-i\theta}\right)\left(q^{2}+re^{\psi+i\theta}\right)dp^{2}\\
 4p^{2}q^{2}re^{-i\theta}dq^{2}+\frac{e^{-\psi}p^{2}r^{2}}{4}\left(q^{2}re^{\psi-i\theta}-1\right) \left(re^{\psi+i\theta}-q^{2}\right)(\psi,_{r})^{2}d\theta^{2}+ipr^{2}\left(q^{2}e^{-i\theta}-e^{i\theta}\right)(\psi,_{r})dpd\theta\\
-ip^{2}qre^{-\psi}\left(1-2q^{2}re^{\psi-i\theta}+e^{2\psi}r^{2}\right)(\psi,_{r})dqd\theta
+2pq\left(e^{-\psi}+2q^{2}re^{-i
\theta}+e^{\psi}r^{2}\right)dqdp\bigg]. \label{PMOne}
\end{multline}

\subsection{Finite gauge group}

In this subsection we work out with hamiltonian vector fields whose
hamiltonian functions correspond to the $\mathfrak{su}(2)$ on
$\mathbb{CP}^1$ with coordinates $(p,q)$. In this case the metrics
are defined on $\mathcal{M}_{1}\times\mathbb{CP}^{1}$, the four-form
is $\nu=dz\wedge d\ove{z}\wedge\Omega_{\mathbb{CP}^{1}}$ and the
vector fields are \cite{Dunajski96}
$$
W=\frac{1}{2}\Phi^{\ast\, i}X_{H_{i}}, \ \ \ \ \ \ \ \ \ \ \ \ \ \ \ Z=\p_{z}+A^{i}_{z}X_{H_{i}}, \no\\
$$
\begin{equation}
\widetilde{W}=\frac{1}{2}\Phi^{i}X_{H_{i}}, \ \ \ \ \ \ \ \ \ \ \ \
\ \ \ \widetilde{Z}=\p_{\ove{z}}+A^{i}_{\ove{z}}X_{H_{i}}.
\end{equation}
The dual frame (\ref{Marco4}) is given by \beq
e_{W}&=&\frac{f^{-2}}{2}\big[\{\Phi^{i}H_{i},A_{z}^{i}H_{i}\}dz+\{\Phi^{i}H_{i},A_{\ove{z}}^{i}H_{i}\}d\ove{z}+d_{\Sigma}(\Phi^{i}H_{i})\big]\no\\
&=&\frac{f^{-2}}{2}\Phi^{i}\big[H_{i,p}dp+H_{i,q}dq-2\varepsilon_{ij}^{k}H_{k}(A_{z}^{j}dz+A_{\ove{z}}^{j}d\ove{z})\big]\no\\
&=&\frac{f^{-2}}{2}\Phi^{i}\left(H_{i,p}D_{p}+H_{i,q}D_{q}\right ), \no\\
e_{\widetilde{W}}&=&\frac{f^{-2}}{2}\big[-\{\Phi^{\ast\, i}H_{i},A_{z}^{i}H_{i}\}dz-\{\Phi^{\ast\, i}H_{i},A_{\ove{z}}^{i}H_{i}\}d\ove{z}-d_{\Sigma}(\Phi^{\ast\, i}H_{i})\big]\no\\
&=&-\frac{f^{-2}}{2}\Phi^{\ast\, i}\big[ H_{i,p}dp+H_{i,q}dq-2\varepsilon_{ij}^{k}H_{k}(A_{z}^{j}dz+A_{\ove{z}}^{j}d\ove{z})\big]\no\\
&=&-\frac{f^{-2}}{2}\Phi^{\ast\, i}\left (H_{i,p}D_{p}+H_{i,q}D_{q}\right ), \no\\
e_{Z}&=&dz, \no\\
e_{\widetilde{Z}}&=&d\ove{z}, \eeq where $Dp$ and $Dq$ are defined
as follows \be Dp:=dp+\frac{A_{\mu}^{i}}{\Omega}H_{i,q}dx^{\mu},
\quad Dq:=dq-\frac{A_{\mu}^{i}}{\Omega}H_{i,p}dx^{\mu}\label{Dq},
\ee with $A_{\mu}dx^{\mu}=A_{z}dz+A_{\ove{z}}d\ove{z}$. Moreover,
the function $f^{2}$ is obtained by \be f^{2}=2\Phi^{\ast\,
i}\Phi^{j}\varepsilon_{ji}^{ \ \ k}h_{k}. \ee

\subsubsection{Liouville equation}

In the case of solutions (\ref{Mosna4}), the hamiltonian vector
fields are given by
$$
D_{z}=\p_{z}-i\p_{z}(\ln g)X_{H_{1}}, \ \ \ \ \ \ \ \ \ \ \ \ \ \frac{1}{2}\Phi^{\ast}=\frac{1}{2}(-gi^{n}X_{H_{2}}-gi^{n+1}X_{H_{3}}), \no\\
$$
\begin{equation}
D_{\ove{z}}=\p_{\ove{z}}+i\p_{\ove{z}}(\ln g)X_{H_{1}}, \ \ \ \ \ \
\ \ \ \ \ \ \
\frac{1}{2}\Phi=\frac{1}{2}(gi^{n}X_{H_{2}}-gi^{n+1}X_{H_{3}}).
\end{equation}
In the present case $Dp$, $Dq$ and $f^{2}$ take the form
\begin{gather}
Dp=dp-i(p^{2}-1)r\p_{r}(\ln g)d\theta, \quad Dq=dq+i(q^{2}-1)r\p_{r}(\ln g)d\theta, \no\\
f^{2}=\pm ig^{2}\frac{p+q}{1+pq}.
\end{gather}
The dual frame is
\begin{eqnarray}
e_{\Phi^{\ast}}&=&\frac{1}{2i^{n}(p+q)(1+pq)g}\left[(1+p)^{2}Dq-(q-1)^{2}Dp\right], \no\\
e_{\Phi}&=&\frac{1}{2i^{n}(p+q)(1+pq)g}\left[-(p-1)^{2}Dq+(q+1)^{2}Dp\right].
\end{eqnarray}

Gathering all that, the metric is a function of $g$, which is
obtained from (\ref{g}), and it has the following form
\begin{multline}
ds^{2}=\pm\frac{i(p+q)g^{2}}{1+pq}dr^{2}+\frac{1}{(p+q)(1+pq)}\bigg\{ir^{2}\big[\pm (p+q)^{2}g^{2}-(p^{2}-1)(q^{2}-1)(\p_{r}(\ln g))^{2}\big]d\theta^{2}\\
+r(q^{2}-1)\big[\p_{r}(\ln g) \big]dpd\theta-r(p^{2}-1)\big[\p_{r}(\ln g)\big]dqd\theta+\frac{i(q^{2}-1)^{2}}{4(1+pq)^{2}}dp^{2}\\
-\frac{i(\big[(1+pq)^{2}+(p+q)^{2}\big]}{2(1+pq)^{2}}dpdq+\frac{i(p^{2}-1)^{2}}{4(1+pq)^{2}}dq^{2}\bigg\}
\end{multline}
If we take $\nu=1$ we get
\begin{multline}
ds^{2}=\pm\frac{4i(p+q)}{(1+pq)(1\mp r^{2})}dr^{2}+\frac{1}{(p+q)(1+pq)}\bigg\{\pm i\left(\frac{p+q}{r^{2}}-(p^{2}-1)(q^{2}-1)\right)M^{2}\\
-(q^{2}-1)Mdp+(p^{2}-1)Mdq+\frac{i(q^{2}-1)^{2}}{4(1+pq)^{2}}dp^{2}-\frac{i\big[(1+pq)^{2}+(p+q)^{2}\big]}{2(1+pq)^{2}}dpdq\\
+\frac{i(p^{2}-1)^{2}}{4(1+pq)^{2}}dq^{2}\bigg\}. \label{LMTwo}
\end{multline}

\subsubsection{Sinh-Gordon equation}

In this case we use the solutions (\ref{Alpha8}) and then
$$
D_{z}=\p_{z}-i\p_{z}(\alpha)X_{H_{1}}, \ \ \ \ \ \
\ \ \ \ \frac{1}{2}\Phi=\frac{\kappa}{2} \cosh(\alpha)X_{H_{2}}-\frac{i\kappa}{2} \sinh(\alpha)X_{H_{3}}, \no\\
$$
\begin{equation}
D_{\ove{z}}=\p_{\ove{z}}+i\p_{\ove{z}}(\alpha)X_{H_{1}}, \ \ \ \ \ \
\ \ \ \ \frac{1}{2}\Phi^{\ast}=-\frac{\kappa}{2}
\cosh(\alpha)X_{H_{2}}+\frac{i\kappa}{2} \sinh(\alpha)X_{H_{3}}.
\end{equation}
Thus we find
\begin{gather}
Dp=dp+\frac{\kappa(p^{2}-1)}{2} \operatorname{csch}(\beta)(e^{-i\omega}dz-e^{i\omega}d\ove{z}),\no\\
Dq=dq-\frac{\kappa(q^{2}-1)}{2} \operatorname{csch}(\beta)(e^{-i\omega}dz-e^{i\omega}d\ove{z}), \no\\
f^{2}=\frac{i\kappa^{2}(p+q)}{1+pq} \operatorname{csch}(\beta)
\operatorname{coth}(\beta).
\end{gather}
The frame is written as \small \beq
e_{\Phi^{\ast}}&=&\frac{\tanh(\beta)}{2\kappa (p+q)(1+pq)}\bigg\{-\big[2p+(1+p^{2})\cosh(\beta)\big]Dq+\big[-2q+(1+q^{2})\cosh(\beta)\big]Dp \bigg\},\no\\
e_{\Phi}&=&\frac{\tanh(\beta)}{2\kappa(p+q)(1+pq)}\bigg\{
\big[-2p+(1+p^{2})\cosh(\beta)\big]Dq-
\big[2q+(1+q^{2})\cosh(\beta)\big]Dp\bigg\}. \eeq \normalsize We
take the simplest case, $\beta=\kappa x$, and it yields the
following metric
\small
\begin{multline}
ds^{2}=\frac{1}{1+pq}\bigg\{i\kappa^{2}(p+q) \coth(\kappa x) \operatorname{csch}(\kappa x)dx^{2}\\
-\frac{i\kappa^{2}\big[1-3q^{2}+p^{2}(q^{2}-3)+(p(q-1)-q-1)(p+q+pq-1)\cosh(2\kappa x)\big] \operatorname{csch}^{2}(\kappa x) \operatorname{sech}(\kappa x)}{2(p+q)}dy^{2}\\
-\frac{\kappa\big[-1+3q^{2}+pq(q^{2}-3)+(pq-1)(q^{2}+1)\cosh(2\kappa x)\big] \operatorname{csch}(2\kappa x)}{(p+q)(1+pq)}dydp\\
+\frac{\kappa\big[(p^{2}+1)(pq-1) \coth(\kappa x)+4p(p-q) \operatorname{csch}(2\kappa x)\big]}{(p+q)(1+pq)}dydp\\
+\frac{i\big[-4q^{2}+(q^{2}+1)^{2} \cosh^{2}(\kappa x)\big] \operatorname{sech}(\kappa x)}{4(p+q)(1+pq)^{2}}dp^{2}\\
-\frac{i(\big[pq+(1+p^{2})(1+q^{2}) \cosh^{2}(\kappa x)\big]\operatorname{sech}(\kappa x)}{2(p+q)(1+pq)^{2}}dpdq\\
+\frac{i\big[-4p^{2}+(p^{2}+1)^{2}\cosh^{2}(\kappa
x)\big]\operatorname{sech}(\kappa
x)}{4(p+q)(1+pq)^{2}}dq^{2}\bigg\}. \label{SGMTwo}
\end{multline}
\normalsize

\subsubsection{Painlev\'e III equation}

In the case of the solutions given by Ward, from (\ref{Ward4}), the
hamiltonian vector fields are given by
$$
D_{z}=\p_{z}-\frac{i}{2}\psi,_{z}X_{H_{3}}, \ \ \ \ \ \ \ \ \ \ \ \ \ \frac{1}{2}\Phi^{\ast}=-\frac{i}{2}(\ove{z}e^{\frac{\psi}{2}}-e^{-\frac{\psi}{2}})X_{H_{1}}-\frac{1}{2}(\ove{z}e^{\frac{\psi}{2}}+e^{-\frac{\psi}{2}})X_{H_{2}}, \no \\
$$
\begin{equation}
D_{\ove{z}}=\p_{\ove{z}}+\frac{i}{2}\psi,_{\ove{z}}X_{H_{3}}, \ \ \
\ \ \ \ \ \ \ \ \ \ \frac{1}{2}\Phi
=-\frac{i}{2}(ze^{\frac{\psi}{2}}-e^{-\frac{\psi}{2}})X_{H_{1}}+\frac{1}{2}(ze^{\frac{\psi}{2}}+e^{-\frac{\psi}{2}})X_{H_{2}}.
\end{equation}
Now the form of $Dp$, $Dq$, $f^{2}$ is given as follows
\begin{gather}
Dp=dp+ipr\psi_{r}d\theta, \quad Dq=dq-iqr\psi_{r}d\theta, \no\\
f^{2}=\frac{i\left(e^{\psi}r^{2}-e^{-\psi}\right)}{2(1+pq)}.
\end{gather}
While the dual frame takes the form \beq
e_{\Phi^{\ast}}&=&\frac{e^{-\frac{\psi}{2}}}{(1+pq)(e^{\psi}r^{2}-e^{-\psi})}\bigg[(p^{2}+re^{i\theta+\psi})Dq-(1+q^{2}re^{i\theta+\psi})Dp\bigg], \no\\
e_{\Phi}&=&\frac{e^{-\frac{\psi}{2}}}{(1+pq)(e^{\psi}r^{2}-e^{-\psi})}\bigg[-(1+p^{2}re^{\psi-i\theta})Dq+(q^{2}+re^{\psi-i\theta})Dp\bigg].
\eeq In this case the metric is given by
\begin{multline}
ds^{2}=\frac{1}{2(pq+1)\left(r^2 e^{\psi}-e^{-\psi}\right)}\bigg\{i\left(r^2 e^{\psi}-e^{-\psi}\right)^{2}dr^2\\
-ir^2 \bigg[(e^{-\psi-i\theta}\left(p r e^{\psi}+q\right) \left(p+q re^{\psi+i\theta}\right)\psi,_{r}^2-\left(r^2 e^{\psi}-e^{-\psi}\right)^2\bigg]d\theta^2\\
+\frac{ie^{-\psi-i \theta}\left(re^{\psi}+e^{i\theta}q^2\right)\left(1+q^2re^{\psi+i\theta}\right)}{(p q+1)^2}dp^2+\frac{ie^{-\psi-i\theta } \left(p^2+r e^{\psi+i\theta}\right)\left(p^2 r e^{\psi}+e^{i \theta }\right)}{(pq+1)^2}dq^2\\
-\frac{r\big[2re^{-i\theta}\left(p+e^{2 i \theta } q^3\right)+q(qp+1)e^{\psi}+qr^2 (pq+1)e^{\psi}\big]\psi,_{r}}{p q+1}dpd\theta\\
+\frac{r\big[2 re^{-i\theta}\left(p^3+e^{2 i \theta }q\right)+p(p q+1)e^{-\psi}+p r^2 (p q+1) e^{\psi}\big]\psi,_{r}}{pq+1}dqd\theta\\
-\frac{i\big[e^{-\psi}\left(p^2 q^2+1\right)+r^2 \left(p^2
q^2+1\right)e^{\psi}+2re^{-i\theta} \left(p^2+e^{2 i \theta
}q^2\right)\big]}{(p q+1)^2}dpdq\bigg\}. \label{PMTwo}
\end{multline}

\section{Conclusions}

In the present paper we briefly survey Hitchin's equations and some
of their solutions. Besides we focused in two types of models
leading to three non-linear equations, namely: Liouville,
sinh-Gordon and Painlev\'e III.

On the other hand, Hitchin's equations constitutes a system of PDE's
among a set of integrable reductions to two dimensions coming from
the SDYM equation in four dimensions. It is well known that some
other reductions to two-dimensional models, such as the WZW model
and the chiral model are strongly related to self-dual gravity in
four dimensions. Unlike these cases, a straightforward relationship
between Hitchin's equations and self-dual gravity is not so evident.
In the present paper we intended to fill this gap. In order to do
that we first follow the strategy employed in several works
\cite{Compean,Plebanski:1996np,Formanski:2004dd,Formanski:2005wt,
GarciaCompean:1996np,GarciaCompean:2009cg}. In this work we first
promote SU$(N)$ Hitchin's equations to a unitary anti-self-dual
operator Lie algebra and then it is taken the gauge fields valued on
this algebra. We used the WWMG-formalism by writing this operator
algebra in a phase-space representation, i.e. in terms of the
coordinate and momentum operators $(\hat{p},\hat{q})$ satisfying the
Heisenberg algebra. In this way we obtained the Moyal deformation of
the SU$(N)$ Hitchin's equations. For the well known relationship
between the Moyal bracket and the Poisson bracket it is possible to
find the large-$N$ limit of the Hitchin's system by taking the limit
$\hbar \to 0$. This is explicitly done for the three cases we
mentioned before: Liouville, Sinh-Gordon and Painlev\'e III. This
limit is characterized by the set of AJS equations for four
hamiltonian vector fields. The volume form of the underlying
spacetime manifold determines the tetrad system and consequently the
half-flat metric.

In section 3, the WWMG-formalism was employed and we found that
SU$(\infty)$ Hitchin's equations (\ref{Hitchingrav4}) and
(\ref{Hitchingrav3}) are related to Husain-Park equation
(\ref{HPE}). Indeed, we see that if we choose a gauge in which
$\Phi=2A_{z}$  and $\Phi^{\ast}=2A_{\ove{z}}$, then the Hitchin's
equations reduce to (\ref{Hitchinredu}) and then, taking the
large-$N$ limit, these equations describe the Husain-Park equation
(\ref{HPE}). In section 5 the half-flat metric was determined for
the mentioned three cases. We found the general metric which is
given by Eq. (\ref{generalHP}). For the specific value of spectral
parameter $\lambda=1$, it reduces to the well known Husein-Park
metric (\ref{specialHP}).  For the three mentioned models the
half-flat metrics are given by Eqs. (\ref{LMOne}), (\ref{SGMOne})
and (\ref{PMOne}) respectively.

In Ref. \cite{Dunajski98} it was alternatively found a description
of a finite dimensional subalgebra of the infinite dimensional one
of vector fields for the gauge theory and its correspondence with
self-dual gravity. For these Hitchin's equations with Lie algebra
$\mathfrak{su}(2)$ we take the underlying manifold as
$\mathbb{C}P^1$ instead of $\mathbb{R}^2$. In section 4 we took this
approach and construct hamiltonian vectors fields satisfying
Hitchin's equations and later in section 5 we use these vectors
fields and found the corresponding self-dual metrics, which for the
three models are given by Eqs. (\ref{LMTwo}), (\ref{SGMTwo}) and
(\ref{PMTwo}) respectively.

There are some questions that we would like to address in the near
future and which we comment briefly in what follows. In Ref.
\cite{Formanski:2005wt} it was found that the Husain-Park equation
can be obtained by dimensional reduction from a master equation and
then a solution, using a Cauchy-Kovalevski form and initial Cauchy
data \cite{Grant:1992ba}, was given. This description is used to
provide an explicit example of a sequence of su$(N)$ chiral fields
giving rise to a curved heavenly space for $N\to\infty$, see also
\cite{Przanowski:1999js}. For that reason, it would be interesting
to determine a Cauchy-Kovalevski form of the SU$(\infty)$ Hitchin's
equations. This can be done by looking for explicit sequences of
su$(N)$ chiral field tending to a curved heavenly space. We will
examine this in a future work.

On the other hand, the Pleba\'nski heavenly equations are the most
usual equations for describing SDG, it is then a natural question to
ask if there exist a relation between SU$(\infty)$ Hitchin's
equations and the heavenly equations. In Ref.
\cite{Jakimowicz:2006gh} Jakimowicz and Tafel shown that Husain-Park
equation is equivalent to the first heavenly equation making a
B\"acklund transformation between these equations. In this direction
it would be interesting to look for a corresponding transformation
between SU$(\infty)$ Hitchin's equations and the heavenly equation.

Finally, as we mentioned in the Introduction, the KW system is a
four-dimensional version of the Hitchin system. These equations are
a dimensional reduction of the Haydys-Witten theory in five
dimensions \cite{khovanov}. Moreover, Haydys-Witten equations are in
turn reductions from the eight-dimensional Spin$(7)$ instanton
equations to five dimensions \cite{Cherkis:2014xua}. In Ref.
\cite{Cardona:2017ebi} it was obtained the SU$(\infty)$
Kapustin-Witten equations and one possible interesting question is
to study what kind of gravity emerges from them as a
higher-dimensional version of the present work.

 \vspace{.5cm}
\centerline{\bf Acknowledgments} \vspace{.5cm}

It is a pleasure to thank Maciej Dunajski and Andr\'es Luna for very
useful comments and suggestions. E. Chac\'on thanks CONACyT and
COMECyT for the support granted.


\newpage


\bibliographystyle{unsrt}

\end{document}